\documentclass[twocolumn]{autart}

\usepackage{amsmath,amssymb}
\usepackage{enumerate}
\usepackage{mydefs}
\usepackage{color}
\usepackage{tikz}
\usepackage{ifthen}
\usetikzlibrary{automata, positioning}
\usepackage[normalem]{ulem}
\newcommand{\liu}[1]{\textcolor{black}{#1}}
\newcommand{\russo}[1]{\textcolor{black}{#1}}

\newcommand{\russonew}[1]{\textcolor{black}{#1}}
\newcommand{\russonewnew}[1]{\textcolor{black}{#1}}

\def\complete{T}

\ifthenelse{\equal{\complete}{T}}{
}{
}
\allowdisplaybreaks

\begin{document}
\begin{frontmatter}
		
		\title{Further characterizations of integral input-to-state stability for hybrid systems}
		
		\author[1]{Shenyu Liu}\ead{shenyuliu@bit.edu.cn},		
		\author[2]{Antonio Russo\corauthref{correspondingauthor}}\ead{antonio.russo1@unicampania.it}
		\corauth[correspondingauthor]{Corresponding author.}\ifthenelse{\equal{\complete}{T}}{}{\thanks{This work was  partially supported by the Italian Ministry of Universities and Research (MUR) under contract FSE-REACT-EU, PON Ricerca e Innovazione 2014-2020 DM1062/2021.}}
		
		\address[1]{School of Automation, Beijing Institute of Technology, 100081, Beijing, China}
		\address[2]{Dipartimento di Ingegneria, Universit\`a degli Studi della Campania \lq\lq L. Vanvitelli'', 81031 Aversa, Italy}   

\begin{abstract}
In this work we present further characterizations of integral input-to-state stability (iISS) for hybrid systems. In particular, the equivalence between \russonew{0-input uniform global asymptotic stability} (\russonew{0-UGAS}) plus uniform bounded energy bounded state (UBEBS) and iISS is examined. 
In order to show this equivalence, some necessary conditions for \russonew{0-UGAS} and UBEBS are \russo{provided}. In addition, a \liu{non-smooth Lyapunov characterization} for hybrid systems is proposed and proven. With the help of the aforementioned equivalence, the combination of local iISS and practical iISS, which are defined in this work, is also shown to be equivalent to iISS under one condition on  the local and practical quantifiers.
\end{abstract}
\end{frontmatter}
 
 \section{Introduction}
For modern control problems, there exists a wide variety of dynamical systems which exhibits both continuous and discrete behaviors and cannot be simply described by differential or difference equations, such as switched systems~\cite{DL:03}, impulsive systems~\cite{VL-DDB-PSS:89}, supervisory control systems~\cite{ASM:97}, sampled-data control systems~\cite{TC-BAF:95} and event-triggered systems~\cite{HeemJoha12}.  These dynamical systems can be best described as the so-called hybrid automata or \emph{hybrid systems}. \russo{The attractiveness of hybrid systems theory is augmented by the possibility of adopting its tools to investigate stability and robustness properties of many applications, such as electrical power converters \cite{Johnson21}, automotive \cite{Balucchi13}, robotics \cite{Manchester11}.} 
The study of hybrid systems is not new (cf. \cite{TAH:96}).
Recently, a framework developed in \cite{RG-RS-AT:12,RGS:21} defines hybrid systems as the composition of four elements: a flow set, a jump set, a differential inclusion regulating the system's continuous dynamics and a difference inclusion regulating the system's discrete dynamics. Such framework not only allows to model a wide range of hybrid systems, but it also paves the way to the study of stability and robustness for such systems. 

Furthermore, stability analysis of dynamical systems in the presence of exogenous inputs has also been an important research topic over the last few decades. One of the well-known achievement is the introduction of \emph{input-to-state stability} (ISS) in \cite{ES:89}, which generalizes the notion of $H_\infty$ stability for nonlinear continuous-time systems. Many applications of ISS in analysis and design of feedback systems have been reported later in~\cite{Sontag2008}. A variant of the ISS notion, called \emph{integral input-to-state stability} (iISS), was introduced in~\cite{ES:98} extending $H_2$ stability to nonlinear systems. The iISS property has been investigated both for continuous-time systems in~\cite{DA-ES-YW:00} and for discrete-time systems in~\cite{DA:99}. 
Regarding the stability analysis for hybrid systems in terms of ISS and iISS properties, while ISS for hybrid systems is well-understood in the work \cite{CC-AT:09}, to the best of the authors' knowledge, except for the work \cite{NN-AK-RG:17} where a Lyapunov characterization of iISS hybrid systems is proposed, the notion of iISS is only studied for some particular types of hybrid systems (e.g., \cite{JM-RG:01} for switched systems, \cite{JH-DL-AT:08} for impulsive systems and \cite{SL-AT-DL:21} for a combination of the two). 

As a result, this theoretical work aims at filling in the gaps in the characterization of iISS for hybrid systems. In particular, inspired by the study of the equivalence between \russonew{\emph{0-input uniform global asymptotic stability}} (\russonew{0-UGAS}) plus \emph{uniform bounded energy bounded state} (UBEBS) and iISS for continuous-time systems in~\cite{DA-ES-YW:00b}, in this work we want to examine the same equivalence for hybrid systems. \liu{Such equivalence has been recently proven for many complex systems, including switched systems \cite{HH-JM:18}, impulsive systems \cite{HaimoManc20}, delay systems \cite{ChailGoks22} and time-varying infinite dimensional systems \cite{MancRojas22}}. Meanwhile, as motivated by our previous observation in~\cite{AR-SL-DL-AC:21}, where it was shown that a switched system is iISS under slow switching if and only if it is both locally iISS and practically iISS under slow switching, we would like to verify if such equivalence also holds for hybrid systems in general. 
In our work, some necessary conditions for \russonew{0-UGAS} and UBEBS of the hybrid systems are \russo{provided} at first. Such necessary conditions provide a non-smooth Lyapunov function for the hybrid system. While the Lyapunov theorem for iISS hybrid systems presented in \cite{NN-AK-RG:17} requires the Lyapunov function to be smooth, and smoothing the obtained non-smooth Lyapunov function turns out to be tedious, in this work we will alternatively show iISS by directly proving a non-smooth Lyapunov characterization. Finally, with the help of the aforementioned equivalence between \russonew{0-UGAS} plus UBEBS and iISS, the combination of local iISS and practical iISS is also shown to be equivalent to iISS under one condition on the local and practical quantifiers.

\ifthenelse{\equal{\complete}{T}}{The rest of the paper is organized as follows. Necessary notations and the background of hybrid systems are introduced in Section~\ref{sec:prelim}, together with some stability properties for hybrid systems and the main result of this paper. Then in Section~\ref{sec:similar}, some necessary conditions for \russonew{0-UGAS} and UBEBS are presented. Section~\ref{sec:non-smooth_V} proposes and proves a non-smooth Lyapunov characterization, and the main result is proven in section~\ref{sec:other}. Finally, Section~\ref{sec:conclusion} concludes our work with some brief discussion on the future research.}{}

\section{Preliminaries}\label{sec:prelim}
In this section, some notations on comparison functions from \cite{HK:02} are presented, together with the necessary background concepts of hybrid systems borrowed from \cite{RG-RS-AT:12,NN-AK-RG:17}. In addition, some stability properties for hybrid systems are formally defined, with which our main result can be stated. 

\subsection{Notations}

Let $\N\!:=\!\{0,1,\cdots\}$ denote the set of natural numbers, $\Rsp\!:=\!(0,\infty)$ denote the set of positive real numbers and $\Rp\!:=\![0,\infty)$ denote the set of non-negative real numbers. 
The standard Euclidean norm is denoted by $|\cdot|$. Given a \russo{nonempty} set $\cA\subset\R^n$, define  $\vert x\vert_\cA:=\inf_{y\in\cA}|x-y|$ for any $x\in\R^n$.  \russo{Given an open set $\cX\subset \R^n$ containing a compact set $\cA$, a function $\omega\colon \cX \to \R_{\geq 0}$ is a proper indicator for $\cA$ on $\cX$ if $\omega$ is continuous, $\omega(x)=0$ if and only if $x\in \cA$, and $\omega(x_i)\to +\infty$ when either $x_i$ tends to the boundary of $\cX$ or $\vert x_i \vert \to +\infty$.} \russo{Also define $\B(\cA,r):=\{x\in\R^n:\omega(x)\leq r\}$ for any $\cA\subset\R^n$ and $r\in\Rp$.}

We say $\alpha\in\mathcal{PD}$ (positive definite) if $\alpha:\Rp\to\Rp$ is continuous, $\alpha(0)=0$ and $\alpha(s)>0$ for all $s>0$. We say $\chi\in\cK$ if $\chi\in \PD$ and it is strictly increasing. Moreover,   $\chi\in\cK_{\infty}$ if $\chi\in\cK$ and $\lim_{s\to\infty}\chi(s)=\infty$. Given a function $\gamma:\Rp\to\Rp$, we say $\gamma\in\cL$ if $\gamma$ is continuous, decreasing and $\lim_{s\to\infty}\gamma(s)=0$. Furthermore, we say $\beta(\cdot,\cdot)\in\KL$ if $\beta:\Rp\times \Rp\to\Rp$ is a function such that $\beta(\cdot,t)\in\cK$ for all $t\in\Rp$ and $\beta(s,\cdot)\in\cL$ for all $s\in\Rp$. We also say a function $\beta(\cdot,\cdot,\cdot)\in\KLL$ if $\beta:\Rp\times \Rp\times \Rp\to\Rp$ is a function such that
$\beta(\cdot,t_1,\cdot)\in\KL$ for all $t_1\in\Rp$ and $\beta(\cdot,\cdot,t_2)\in\KL$ for all $t_2\in\Rp$.

\subsection{Hybrid systems}
Consider a hybrid system  $\cH:=\cH(\cC,\cD,\cF,\cG)$ with state $x \in \cX \subset \R^n$ and input $u \in \cU \subset \R^m$
\begin{equation}\label{def:hybrid_sys}
\cH:=\begin{cases}
    \dot x\in\cF(x,u)&(x,u)\in \cC,\\
    x^+\in\cG(x,u)&(x,u)\in \cD.
\end{cases}
\end{equation}
The \emph{flow} and \emph{jump} sets are denoted by $\cC$ and $\cD$, respectively.
Basic regularity assumptions borrowed from \cite{CC-AT:09} are imposed for the hybrid system $\cH$:
\begin{enumerate}[label=A\arabic*]
    \item the set $\cX$ is open, $\cU$ is closed and $\mathcal{C}$ and $\mathcal{D}$ are relatively closed sets in $\cX\times\cU$. \label{Ass:A1}
    \item The set-valued map $\cF\colon \cC\rightrightarrows\R^n$ is outer semi-continuous and locally bounded, and $\cF(x,u)$ is nonempty and convex for all $(x,u)\in\cC$. \label{Ass:A2}
    \item The set-valued map $\cG: \cD\rightrightarrows\cX$ is outer semi-continuous and locally bounded, and $\cG(x,u)$ is nonempty for all $(x,u)\in\cD$. \label{Ass:A3}
\end{enumerate}
We refer to the assumptions \ref{Ass:A1}--\ref{Ass:A3} as \emph{Standing Assumptions}. 
The following definitions are needed in the sequel. A subset $E\subset \Rp\times \N$ is called a \emph{compact hybrid time domain} if $E=\bigcup_{j=0}^J([t_j,t_{j+1}],j)$ for some finite sequence of real numbers $0=t_0\leq \cdots \leq t_{J+1}$. We say $E$ is a \emph{hybrid time domain} if, for each pair $(T,J)\in E$, the set $E\cap([0,T]\times\{0,1,\cdots,J\})$ is a compact hybrid time domain. For each hybrid time domain $E$, there is a natural ordering of points: given $(t_1,j_1),(t_2,j_2)\in E$, $(t_1,j_1)\leq(t_2,j_2)$ (resp. $(t_1,j_1)<(t_2,j_2)$) if $t_1+j_1\leq t_2+j_2$ (resp.  $t_1+j_1< t_2+j_2$). For a hybrid time domain $E$, define
\begin{align*}
    &\sup_t E:=\sup\{t\in\Rp:\exists j\in \N\mbox{ such that }(t,j)\in E\},\\
    &\sup_j E:=\sup\{j\in \N:\exists t\in\Rp\mbox{ such that }(t,j)\in E\},\\
    &\length(E):=\sup_tE+\sup_jE.
\end{align*}
The operations $\sup_t$ and $\sup_j$ on a hybrid time domain $E$ return the supremum of the $\Rp$ and $\N$ coordinates, respectively, of points in $E$. Given a hybrid time domain $E$, a function $x:E\to \cX$ is called a \emph{hybrid signal} and we denote $\dom x:=E$ to be the domain of this hybrid signal. Given a hybrid signal $x:\dom x\to \cX$ and any \russonew{$s \in [0, \sup_t \dom x] \cap \mathbb{R}_{\geq 0}$}, denote $i(s):=\max\{i\in\N:(s,i)\in\dom x\}$. A hybrid signal $x:\dom x\to \cX$ is a \emph{hybrid arc} if for each $j\in\N$, the function \russonew{$t \mapsto x(t,j)$} is locally absolutely continuous on the interval $I^j:=\{t:(t,j)\in\dom x\}$. A hybrid signal $u:\dom u\to \cU$ is a \emph{hybrid input} if for each $j\in\N$, $u(\cdot,j)$ is Lebesgue measurable and locally essentially bounded. 

Let a hybrid input $v:\dom v\to \cU$ be given. Let $(0,0)$,$(t,j)\in\dom v$ be such that $(0,0)<(t,j)$. Define as
\[
\Gamma(v):=\{(t,j)\in\dom v:(t,j+1)\in\dom v\},
\]
the collection of hybrid time of jumps. 
For any $\gamma\in\cK$, define
\begin{equation*}
\Vert v\Vert_{(t,j)}^{\gamma}:=\int_0^t\gamma\big(|v(s,i(s))|\big)\mathrm{d}s+\hspace{-1cm}\sum_{\tiny\begin{array}{c}
     (t',j')\in\Gamma(v),  \\
     (0,0)\leq(t',j')\leq(t,j)
\end{array}}\hspace{-1cm}\gamma\big(|v(t',j')|\big).
\end{equation*}
For any $l\in\Rsp$, we denote the set of all hybrid inputs with $\Vert v\Vert_{(t,j)}^{\gamma}\leq l$ for all $(t,j)\in\dom v$ by $\cL_{\gamma}(l)$. We also denote the set of all hybrid inputs with $\Vert v\Vert_{(t,j)}^{\gamma}<\infty$ for all $(t,j)\in\dom v$ by $\cL_{\gamma}^e$. 

A hybrid arc $x:\dom x\to \cX$ and a hybrid input $u:\dom u\to \cU$ is a \emph{solution pair} $(x,u)$ to $\cH$ if $\dom x=\dom u$, $x(0,0),u(0,0)\in \cC\cup \cD$, and
\begin{itemize}
    \item for each $j\in\N$, $(x(t,j),u(t,j))\in\cC$ and $\dot x\in \cF(x(t,j),u(t,j))$ for almost all $t\in I^j$ where $I^j$ has nonempty interior;
    \item for all $(t,j)\in\Gamma(x)$, $(x(t,j),u(t,j))\in\cD$ and $x(t,j+1)\in\cG((x(t,j),u(t,j)))$.
\end{itemize}
A solution pair $(x,u)$ to $\cH$ is \emph{maximal} if it cannot be extended, it is \emph{complete} if $\length(\dom x)=\infty$. A maximal solution to $\cH$ with the initial condition $x(0,0)=x_0$ and the input $u$ is denoted by $x(\cdot,\cdot;x_0,u)$ and when it is clear from the context, the solution at hybrid time $(t,j)$ is abbreviated by $x(t,j)$. The set of all maximal solution pairs $(x,u)$ to $\cH$ with initial condition $x_0$ \russonew{and input $u\in\cL_\gamma^e$} is designated by $\mathfrak s_u(x_0)$. \russonew{For any compact set $K$, if the set $\bigcup_{x_0\in K}\mathfrak s_u(x_0)$} contains only bounded or complete solution pairs, then we call $\cH$ \emph{pre-forward complete}. 

\subsection{Stability definitions}
\russo{In what follows, let $\cA\subset \cX$ be a nonempty and compact set and $\omega$ be a proper indicator for $\cA$ on $\cX$}. Here we provide some stability definitions with respect to the set $\cA$, which will be discussed in this work.

\begin{defn}\label{def:iISS}
A hybrid system $\cH$ is \emph{integral-input-to-state stable} or iISS \russo{with respect to $\cA$} if there exist functions $\beta\in\KLL$, $\chi\in\cK_{\infty}$ and $\gamma\in\cK$ such that for all $x_0\in\cX,u \in \cL_{\gamma}^e$, each solution pair $(x,u)\in\mathfrak s_u(x_0)$ satisfies
\begin{equation}\label{eqn:iISS}
    \omega(x(t,j;x_0,u))\leq \beta(\omega(x_0),t,j)+ \chi(\Vert u\Vert^{\gamma}_{(t,j)})
\end{equation}
for all $(t,j)\in\dom x$.
\end{defn}

We remark here that the above stability property is called ``pre-iISS" in \cite{NN-AK-RG:17}, where the prefix ``pre-" addresses the cases where not all maximal solutions of $\cH$ are complete. Such technical consideration is also taken into account in this work, as we only require the estimate \eqref{eqn:iISS} to hold for all $(t,j)\in\dom x$. Nevertheless, we omit the prefix ``pre-" here and in the subsequent stability definitions for the sake of presentation clarity.

\begin{defn}\label{def:0-UGAS}
	A hybrid system $\cH$ is \russonew{\emph{0-input uniformly globally asymptotically stable}} or \russonew{0-UGAS} \russo{with respect to $\cA$}, if there exist functions $\beta\in\KLL$ such that for all $x_0\in\cX$, each solution pair $(x,0)\in\mathfrak s_u(x_0)$ satisfies
	\begin{equation}\label{eqn:0-UGAS}
		\russo{\omega(x(t,j;x_0,0))\leq \beta(\omega(x_0),t,j)}
	\end{equation}		
	for all $(t,j)\in\dom x$.
\end{defn}
In Definition~\ref{def:0-UGAS}, the term ``global" is used to indicate that the property holds for all $x_0\in\cX$, even though it is possible that $\cX\neq\R^n$.

\begin{defn}
	A hybrid system $\cH$ is \emph{uniform bounded energy bounded state} or UBEBS \russo{with respect to $\cA$}, if there exist functions $ \alpha, \,\kappa\in\cK_\infty,\gamma\in\cK$ and $c\in\Rsp$ such that for all $x_0\in\cX,u \in \cL_\gamma^e$, each solution pair $(x,u)\in\mathfrak s_u(x_0)$ satisfies
	\begin{equation}\label{eqn:UBEBS}
		\alpha(\omega(x(t,j;x_0,u)))\leq \kappa(\omega(x_0))+ \Vert u\Vert^{\gamma}_{(t,j)}+c
	\end{equation}
	for all $(t,j)\in\dom x$.
\end{defn}

The next definition is a ``local" version of Definition~\ref{def:iISS}.

\begin{defn}\label{def:local iISS}
Let $l\in\Rsp$. A hybrid system $\cH$ is $l$-\emph{locally iISS} \russo{with respect to $\cA$} if there exist functions $\beta\in\KLL$, $\chi\in\cK_{\infty}$ and $\gamma\in\cK$ such that for all $x_0\in \B(\cA,l), u \in \cL_{\gamma}(l)$, each solution pair $(x,u)\in\mathfrak s_u(x_0)$ satisfies \eqref{eqn:iISS} for all $(t,j)\in\dom x$. 
\end{defn}
We also say that a hybrid system $\cH$ is locally iISS if there exists $l\in\Rsp$ such that the system is $l$-locally iISS.

\begin{defn}\label{def:practical iISS}
Let $p\in\Rp$. A hybrid system $\cH$ is $p$-\emph{practically iISS} \russo{with respect to $\cA$} if there exist functions $\beta\in\KLL$, $\chi\in\cK_{\infty}$ and $\gamma\in\cK$ such that for all $x_0\in \cX, u \in \cL_{\gamma}^e$, each solution pair $(x,u)\in\mathfrak s_u(x_0)$ satisfies 
\begin{equation}\label{eqn:practical_iISS}
    \omega(x(t,j;x_0,u))\leq \beta(\omega(x_0),t,j)+ \chi(\Vert u\Vert^{\gamma}_{(t,j)})+p
\end{equation}
for all $(t,j)\in\dom x$. 
\end{defn}
Similarly to Definition~\ref{def:local iISS}, we say that a hybrid system $\cH$ is practically iISS if there exists $p\in\Rp$ such that the system is $p$-practically iISS. Note that iISS is equivalent to $0$-practically iISS. 

We can now state the main result of our work.

\begin{thm}\label{thm:main}
	Consider a hybrid system $\cH$ and suppose that the Standing Assumptions hold. The following facts are equivalent:
	\begin{enumerate}[label=\arabic*)]
		\item The system is iISS.
		\item The system is \russonew{0-UGAS} and UBEBS.
		\item The system is $l$-locally iISS and $p$-practically iISS with $l>p$.
	\end{enumerate}
\end{thm}

It is obvious that $1) \Rightarrow 2)$ and $1) \Rightarrow 3)$. Hence for proving Theorem~\ref{thm:main}, it suffices to show the implications $2) \Rightarrow 1)$ and $3) \Rightarrow 2)$, which are non-trivial and require some intermediate results. In Section~\ref{sec:similar}, we provide some necessary conditions of \russonew{0-UGAS} and UBEBS. Based on these necessary conditions, a non-smooth Lyapunov function can be constructed. Then, in Section~\ref{sec:non-smooth_V}, the aforementioned non-smooth Lyapunov function is shown to imply iISS for the hybrid system. With these intermediate results, our main result is finally proven in Section~\ref{sec:other}. We also give some discussion in Section~\ref{sec:other} on the relation of the quantifiers $l,p$ in $3)$ of Theorem~\ref{thm:main}. 

We remark here that the relation on the quantifiers in $3)$ of Theorem~\ref{thm:main} is critical; when $l\leq p$, $l$-local iISS plus $p$-practical iISS may not imply iISS. To see this, consider the simple continuous-time one dimensional example:
\begin{equation}\label{sys:bad}
	\dot x= -x(x-1)^2+u.
\end{equation}
Note that when $u=0$, $x=0$ is a stable equilibrium for \eqref{sys:bad} while $x=1$ is an unstable equilibrium. In addition $x=1$ serves as a ``separating orbit" such that the solutions with initial state $x_0>1$ will converge to $1$, and solutions with initial state $x_0<1$ will converge to $0$. Therefore, this system is not \russonew{0-UGAS} with respect to the origin and hence not iISS with respect to the origin. 
However, the hybrid system \eqref{sys:bad} can be shown to be both $l$-local iISS and $p$-practically iISS with respect to the origin with any $l<1\leq p$.

\ifthenelse{\equal{\complete}{T}}{
To show practical iISS, define $V_1(x):=\omega(x)^2$ where 
\begin{equation*}
    \omega(x)\coloneqq \begin{cases}
        0 & \text{if} \enskip |x|< 1, \\
        |x|-1 & \text{otherwise},  
    \end{cases}
\end{equation*}
and pick arbitrary $p\geq 1$. Then, for all $x$ with $|x|\geq 1$, we have $\omega(x)+1= |x|$, and 
\begin{align*}
    \dot V_1(x)&=-2x^2(1-x)^2+2xu\\
    &\leq -2|x|^2(1-|x|)^2+2|x||u|\\
    &=-2(\omega(x)+1)^2\omega(x)^2+2(\omega(x)+1)|u|\\
    &\leq -2\omega(x)^4-4\omega(x)^3-\omega(x)^2+|u|^2+2|u|.
\end{align*}
Hence
\begin{equation}\label{iISS_V}
    \dot V_1(x)\leq -\alpha(\omega(x))+\sigma(|u|),
\end{equation}
where $\alpha(s):=2s^4+4s^3+s^2\in\cK_{\infty}$ and $\sigma(s):=s^2+2s\in\cK_{\infty}$. The inequality~\eqref{iISS_V} also holds for all $x$ with $|x|<1$, since $\omega(x)=0$ and $\dot V_1(x)\leq 2|x||u|\leq 2|u|\leq -\alpha(0)+\sigma(|u|)$. Thus by \cite[Theorem 1]{DA-ES-YW:00}, the system \eqref{sys:bad} is iISS with respect to the interval $[-1,\,1]$, which implies the existence of  $\tilde\beta\in\KL,\gamma\in\cK$ such that
\begin{align*}
    |x(t)|&\leq \omega(x)+1\\
    &\leq \tilde\beta(\omega(x_0),t)+\int_0^t\gamma(|u(\tau)|)\mathrm{d}\tau+1\\
    &\leq \tilde\beta(\max\{|x_0|-1,0\},t)+\int_0^t\gamma(|u(\tau)|)\mathrm{d}\tau+1.\\
    &\leq\beta(|x_0|,t)+\int_0^t\gamma(|u(\tau)|)\mathrm{d}\tau+p,
\end{align*}
where $\beta(s,t):=\tilde\beta(\max\{s-1,0\},t)+\frac{s}{t}\in\KL$. Therefore the system \eqref{sys:bad} is $p$-practically iISS.

To show local iISS, we let $V_2(x):=x^2$ and pick arbitrary $l<1$. Define $\gamma(r):=\frac{4l^2r}{1-l^2}$ for all $r\in\Rp$.  We first claim that $|x(t)|\leq 1$ for all $t\geq 0$, all $|x_0|\leq l,\int_0^\infty\gamma(|u(\tau)|) \mathrm{d}\tau\leq l$. If this is not true, denote $T:=\inf\{t\in\Rp:|x(t)|\geq 1\}<\infty$. However, since $\dot V_2(x(t))\leq 2|x(t)||u(t)|$, we have
\begin{align*}
    V_2(x(T))&\leq V(x_0)+2\int_0^T|x(\tau)||u(\tau)|\mathrm{d}\tau\\
    &\leq l^2+2l\int_0^\infty\frac{1-l^2}{4l^2}\gamma(|u(\tau)|)\mathrm{d}\tau\\
    &\leq l^2+\frac{1-l^2}{2}=\frac{l^2+1}{2}<1,
\end{align*}
which is a contradiction. We can in fact conclude that $V_2(x(t))\leq \frac{l^2+1}{2}$ for all $t\in\Rp$. Denote $r:=\sqrt{\frac{l^2+1}{2}}<1$. For all $|x_0|\leq l,\int_0^\infty\gamma(|u(\tau)|) \mathrm{d}\tau\leq l$ and $t\in\Rp$, since $|x(t)|=\sqrt{V_2(x(t))}\leq r$, we conclude that $\dot V_2(x)\leq -2V_2(x)(1-r)^2+2r|u|$. Recall the definition of $V(x)$ and we can conclude from the comparison principle that 
\begin{align*}
    |x(t)|&\leq \sqrt{|x_0|^2 e^{-2(1-r)^2t}+\int_0^t 2re^{-2(1-r)^2(t-\tau)}|u(\tau)|\mathrm{d}\tau} \\
    &\leq \sqrt{2}e^{-(1-r)^2t}|x_0|+\sqrt{\int_0^t 4re^{-2(1-r)^2(t-\tau)}|u(\tau)|\mathrm{d}\tau}, \\
    &\leq  \sqrt{2}e^{-(1-r)^2t}|x_0|+\sqrt{\int_0^t 4r|u(\tau)|\mathrm{d}\tau}, \\
    & = \sqrt{2}e^{-(1-r)^2t}|x_0|+\frac{\sqrt{1-l^2}}{l}\sqrt{\int_0^t \gamma(|u(\tau)|)\mathrm{d}\tau}.
\end{align*}
Therefore we conclude \eqref{eqn:iISS} with $\beta(s,t) := \sqrt{2}e^{-(1-r)^2t}s$, $\chi(s):=\frac{\sqrt{1-l^2}}{l}\sqrt{s}$ and $\gamma(s)$ defined as earlier, which implies that the system \eqref{sys:bad} is $l$-locally iISS.}{The detailed computation can be found in \cite{SL-AR:22-arxiv}.}

\section{Necessary conditions for \russonew{0-UGAS} and UBEBS}\label{sec:similar}
In this section, some necessary conditions for \russonew{0-UGAS} and UBEBS hybrid systems will be studied. 

\subsection{Necessary conditions for \russonew{0-UGAS}}\label{subsec:0-UGAS}
The first lemma is a consequence of \cite[Lemma 6]{NN-AK-RG:17}.

\begin{lem}\label{lem:0-UGAS_to_Lyap_char}
    \liu{Let $\cA\subset\cX$ be a nonempty compact set.} A hybrid system $\cH$ is \russonew{0-UGAS} if and only if there exist a smooth Lyapunov function $V\colon \cX \to \mathbb{R}_{\geq 0}$, functions $\alpha_1$, $\alpha_2$, $\alpha_3$, \russo{$\alpha_4$} $\in \mathcal{K}_{\infty}$ and a function $q \colon \mathbb{R}_{\geq 0}\to \mathbb{R}_{> 0}$ with the property that $q(s)= 1$ for all $s\in[0,1]$ such that
\begin{align}
       \alpha_1(\omega(x)) \leq V(x)&\leq \alpha_2(\omega(x)) \quad \forall x\in \cX, \label{eq:lemmaUBEBS_1} \\ 
        \liu{\max_{f\in \cF(x,q(\omega(x))\nu)}\langle\nabla V(x),f \rangle}&\liu{\leq -\alpha_3(\omega(x)) + \alpha_4(|\nu|)} \nonumber \\ & \quad   \liu{\forall (x,q(\omega(x))\nu)\in\cC},  \label{eq:lemmaUBEBS_equiV_c}\\
	    \liu{\max_{g\in \cG(x,q(\omega(x))\nu)}V(g)-V(x)}&\liu{\leq -\alpha_3(\omega(x)) + \alpha_4(|\nu|)} \nonumber \\ & \quad \liu{\forall (x,q(\omega(x))\nu)\in\cD}. \label{eq:lemmaUBEBS_equiV_d}    
\end{align}

\end{lem}

\begin{pf}
    The ``if" part is clear to show, by setting $\nu=0$ in \eqref{eq:lemmaUBEBS_equiV_c} and \eqref{eq:lemmaUBEBS_equiV_d}. To show the ``only if" part, we appeal to \cite[Lemma 6]{NN-AK-RG:17}, which can be easily modified and proven for the case of differential/difference inclusions such that, $\cH$ being \russonew{0-UGAS} implies the existence of a smooth Lyapunov function $V\colon \mathcal{X} \to \mathbb{R}_{\geq 0}$, functions $\alpha_1,\alpha_2,\hat\alpha_3, \chi \in \mathcal{K}_{\infty}$ and a smooth function $q \colon \mathbb{R}_{\geq 0}\to \mathbb{R}_{> 0}$ with the property that $q(s)= 1$ for all $s\in[0,1]$ \russo{($q(s)$ constructed as in \cite[Lemma 3.1]{Sontag90})} such that, \russo{condition} \eqref{eq:lemmaUBEBS_1} holds together with
    \begin{align}
        \max_{f\in \cF(x,q(\omega(x))\nu)}\langle\nabla V(x),f \rangle&\leq -\hat\alpha_3(\omega(x)) \nonumber \\ &\hspace{-3cm} \quad \forall (x,q(\omega(x))\nu)\in\cC \enskip \text{with} \enskip \omega(x)\geq\chi(\rvert\nu \lvert),  \label{eq:lemmaUBEBS_2} \\
 		\max_{g\in \cG(x,q(\omega(x))\nu)}V(g)-V(x)&\leq -\hat\alpha_3(\omega(x)) \nonumber \\ &\hspace{-3cm}\quad\forall (x,q(\omega(x))\nu)\in\cD  \enskip \text{with} \enskip \omega(x)\geq\chi(\rvert\nu \lvert). \label{eq:lemmaUBEBS_3}
	 \end{align}
    To prove the implication from \eqref{eq:lemmaUBEBS_2} to \eqref{eq:lemmaUBEBS_equiV_c}, define $\alpha_{4c}(r):=\max\{0,\hat\alpha_{4c}(r)\}$, where 
    \begin{equation*}
        \hat\alpha_{4c}(r)\!:=\hspace{-.5cm}\max_{\tiny\begin{array}{c}
             |\nu| \leq r, \\
             \omega(x)\leq \chi(\russo{|\nu|})
        \end{array}}\hspace{-.5cm}\left(\max_{f\in \cF(x,q(\omega(x))\nu)}\!\!\langle\nabla V(x),f \rangle  +\hat\alpha_3(\omega(x))\!\right)\!.
    \end{equation*}
    \liu{Clearly, by its definition, $\alpha_{4c}$ is non-decreasing and $\alpha_{4c}(0)=0$, since \eqref{eq:lemmaUBEBS_2} implies $\max_{f\in\cF(x,0)}\langle\nabla V(x),f\rangle\leq 0$ for all $x\in\cA$. Suppose $\alpha_{4c}(r)$ is not continuous at $0$, then there exists $\delta>0$, $x^*\in\cA$, a convergent sequence of $(x_i,\nu_i)\to (x^*,0)$ satisfying $\omega(x_i)\leq \chi(|\nu_i|)$ such that $\limsup_{i\to\infty}\max M(x_i,\nu_i)\geq\delta$, where $M(x_i,\nu_i):=\{\langle\nabla V(x_i),f \rangle  +\hat\alpha_3(\omega(x_i)):f\in \cF(x_i,q(\omega(x_i))\nu_i)\}\subset\R$. On the other hand, it follows from the outer semi-continuity of $\cF$ that $\limsup_{i\to\infty} M(x_i,\nu_i)\subset M(x^*,0)\subset(-\infty,0]$, which is a contradiction. Therefore, $\alpha_{4c}$ is 0 at 0, continuous at 0 and non-decreasing, which implies that} it can be majorized to be a class $\cK_{\infty}$ function. 
    Analogously, we can construct $\alpha_{4d}\in\cK_\infty$. Define $\alpha_3(r):=\hat\alpha_3(r)$ and $\alpha_4(r):=\max\{\alpha_{4c}(r),\alpha_{4d}(r)\}$ for all $r\in\Rp$. Note that if $\omega(x)\geq \chi(|\nu|)$, then \eqref{eq:lemmaUBEBS_2} directly implies \eqref{eq:lemmaUBEBS_equiV_c}; otherwise if $\omega(x)\leq \chi(|\nu|)$, then \eqref{eq:lemmaUBEBS_equiV_c} still holds because $\alpha_4(r)\geq \sup_{|\nu|=r} \max_{f\in \cF(x,q(\omega(x))\nu)}\langle\nabla V(x),f \rangle  +\alpha_3(\omega(x))$. 
    Analogously, we can also conclude the implication from \eqref{eq:lemmaUBEBS_3} to \eqref{eq:lemmaUBEBS_equiV_d}.
    
\end{pf}
\begin{rem}
    We point out here that the proof in \cite[Lemma 6]{NN-AK-RG:17} relies on the Converse Lyapunov Theorem in \cite{CC-AT-RG:08}, which is guaranteed by our Standing Assumptions. \russo{Furthermore, Lemma \ref{lem:0-UGAS_to_Lyap_char} is reminiscent of \cite[Lemma IV.10]{DA-ES-YW:00}, whose proof, however, is based on the presence of a single isolated equilibrium point and cannot thus be straightforwardly inherited to analyze stability properties of hybrid systems with respect to compact sets.}
\end{rem}

We further provide two necessary conditions of \russonew{0-UGAS} based on Lemma~\ref{lem:0-UGAS_to_Lyap_char}. The first one, proven below, states that \russonew{0-UGAS} implies local iISS.
\begin{lem}\label{lem:0GAS_to_LiISS}
     \liu{Let $\cA\subset\cX$ be a nonempty compact set.} If a hybrid system $\cH$ is \russonew{0-UGAS} with respect to $\cA$, then it is also locally iISS with respect to $\cA$. 
\end{lem}
\begin{pf}
Recalling Lemma~\ref{lem:0-UGAS_to_Lyap_char}, we claim that the inequalities \eqref{eq:lemmaUBEBS_1}, \eqref{eq:lemmaUBEBS_equiV_c} and \eqref{eq:lemmaUBEBS_equiV_d}  imply $l$-local iISS for $\cH$ with $l$ satisfying 
	    \begin{equation}\label{eq:l_condition}
	        \alpha_2(l)+l\leq \alpha_1(1).
	    \end{equation}
	    In fact, because $q(s)= 1$ for all $s\in[0,1]$, it follows from \eqref{eq:lemmaUBEBS_equiV_c} and \eqref{eq:lemmaUBEBS_equiV_d} that
	    \begin{align}
	        \max_{f\in\cF(x,u)}\langle\nabla V(x),f \rangle&\leq -\alpha_3(\omega(x)) + \gamma(|u|)\nonumber \\ 
	        &\hspace{-0.5cm} \forall (x,u)\in\cC \enskip \text{with} \enskip x\in\B(\cA,1), \label{eq:lemmaUBEBS_6}  \\
	 		\max_{g \in \cG(x,u)} V(g)-V(x)&\leq -\alpha_3(\omega(x)) + \gamma(|u|)\nonumber \\ 
	 		& \hspace{-0.5cm} \forall (x,u)\in\cD  \enskip \text{with} \enskip x\in\B(\cA,1), \label{eq:lemmaUBEBS_7} 
	    \end{align}
	    where $\gamma:=\alpha_4\in\cK_\infty$.
	    
	    Let $z(t,j):=\Vert u\Vert^{\gamma}_{(t,j)}$. We have
    \begin{align*}
        \dot z(t,j)&=\gamma(|u(t,j)|),\quad (t,j)\in \dom x\backslash\Gamma(x),\\
        z(t,j+1)-z(t,j)&=\gamma(|u(t,j)|),\quad (t,j)\in\Gamma(x).
    \end{align*}
    By definition, $z(t,j)$ is non-negative and non-decreasing and when $u\in\cL_{\gamma}(l)$, $z(t,j)\leq l$ for all $(t,j)\in\dom x$. Further define $y(t,j):=V(x(t,j))-z(t,j)$. 
    We now claim that the set $I:=\{y\in\R:y\leq \alpha_2(l)\}$ is forward-invariant for $y(\cdot,\cdot)$. We prove the claim by contradiction. Suppose $I$ is not forward-invariant, then there exists a hybrid time $(t^*,j^*)\in\dom x$ such that either $(t^*,j^*)\in\dom x\backslash\Gamma(x)$ , $y(t^*,j^*)=\alpha_2(l)$ (which also implies $\omega(x(t^*,j^*))\geq \alpha_2^{-1}(V(x(t^*,j^*)))\geq \alpha_2^{-1}(y(t^*,j^*))=l$) and $\dot y(t^*,j^*)\geq 0$, or $(t^*,j^*)\in\Gamma(x)$, $y(t^*,j^*)\leq \alpha_2(l)$ but $y(t^*,j^*+1)> \alpha_2(l)$. Since in either case, $y(t^*,j^*)\in I$, it is implied by \eqref{eq:l_condition} that
    \begin{align}
    \omega(x(t^*,j^*))&\leq \alpha_1^{-1}(V(x(t^*,j^*))) \nonumber \\
    &= \alpha_1^{-1}(y(t^*,j^*)+z(t^*,j^*))\nonumber \\
    &\leq \alpha_1^{-1}(\alpha_2(l)+l)\leq 1.    \label{bounding_x(t,j)}
    \end{align}
    Thus \eqref{eq:lemmaUBEBS_6},\eqref{eq:lemmaUBEBS_7} are applicable and we conclude that either $(t^*,j^*)\in\dom x\backslash\Gamma(x)$ and
    \begin{align}
        \dot y(t^*,j^*)&\leq \max_{f\in\cF(x,u)}\langle\nabla V(x),f \rangle-\dot z(t^*,j^*) \nonumber \\
        &\leq -\alpha_3(\omega(x(t^*,j^*))) \leq -\alpha_3(l)<0,\label{LiISPS_step_c1}
        \end{align}
        or $(t^*,j^*)\in\Gamma(x)$ and 
        \begin{align}
        y&(t^*,j^*+1)-y(t^*,j^*) \nonumber \\
        &\leq \big(\max_{g \in \cG(x,u)} V(g)-V(x)\big)-\big(z(t^*,j^*+1)-z(t^*,j^*)\big)\nonumber \\
        &\leq -\alpha_3(\omega(x(t^*,j^*)))\leq 0,\label{LiISPS_step_d1}
    \end{align}
    which are contradictions. Hence $y(t,j)$ can not escape $I$ neither by continuous flow nor by a discrete jump and this completes the proof for the claim; $I$ is forward invariant. Note that when $x_0\in\B(\cA,l)$, $y(0,0)=V(x_0)\leq \alpha_2(l)$ so $y(0,0)\in I$ and in fact $y(t,j)\in I$ for all $(t,j)\in\dom x$. Furthermore, by a similar argument as \eqref{bounding_x(t,j)}, we can conclude that $x(t,j)\in\B(\cA,1)$ so \eqref{eq:lemmaUBEBS_6}, \eqref{eq:lemmaUBEBS_7} are applicable for all $(t,j)\in\dom x$. By a similar derivation as \eqref{LiISPS_step_c1}, \eqref{LiISPS_step_d1}, we then conclude that
    \begin{align*}
        \dot y(t,j)
        \!&\leq\! -\alpha_3(\omega(x(t,j)))\enskip\forall (t,j)\in\dom x\backslash\Gamma(x), \\
        y(t,j\!+\!1)-y(t,j)\!&\leq\! -\alpha_3(\omega(x(t,j)))\enskip\forall (t,j)\in\Gamma(x),
    \end{align*}
    which, after exploiting \eqref{eq:lemmaUBEBS_1}, gives
    \begin{equation*}
        \dot y(t,j)\leq  -\alpha_3\circ\alpha_2^{-1}(\max\{y(t,j)+z(t,j),0\})
        \end{equation*}
        for all $(t,j)\in\dom x\backslash\Gamma(x)$ and 
        \begin{equation*}
        y(t,j+1)-y(t,j) \leq -\alpha_3\circ\alpha_2^{-1}(\max\{y(t,j)+z(t,j),0\})
    \end{equation*}
    for all $(t,j)\in\Gamma(x)$. Then by \cite[Lemma 9]{NN-AK-RG:17}, there exists $\tilde\beta\in\KLL$ such that
    \begin{equation*}\label{second_part_i}
    y(t,j)\leq \max\{\tilde\beta(y(0,0),t,j),z(t,j)\}\quad \forall (t,j)\in \dom x.  
    \end{equation*}
    Finally, it follows from \eqref{eq:lemmaUBEBS_1} that when $x_0\in\B(\cA,l)$ and $u\in\cL_\gamma(l)$, 
    \begin{align*}
    \alpha_1(\omega(x(t,j)))&\leq V(x(t,j)) = y(t,j)+z(t,j)  \\
    &\leq \tilde\beta(y(0,0),t,j)+2\Vert u\Vert^{\gamma}_{(t,j)} \\ 
    &\leq \tilde\beta(\alpha_2(\omega(x_0)),t,j)+2\Vert u\Vert^{\gamma}_{(t,j)},
\end{align*}
for all $(t,j)\in \dom x$. Thus, \liu{applying $\alpha_1^{-1}$ on both sides of the above inequality, \eqref{eqn:iISS} is proven with  $\beta(s,t,j):=\alpha_1^{-1}\left(2\tilde\beta(\alpha_2(s),t,j)\right)$, $\chi(s)=\alpha_1^{-1}\left(4s\right)$.}\hfill$\blacksquare$
\end{pf}


To state the second necessary condition of \russonew{0-UGAS}, we need to introduce the notion of \emph{semiproper} functions.
Given a set $\cA\subset\cX$, a function $W:\cX\to\Rp$ is semiproper if there exist a function $\pi\in\cK$ and a continuous function $W_0:\cX\to\Rp$ with the properties that $W_0(x)=0$ for all $x\in\cA$, $W_0(x)>0$ for all $x\not\in\cA$ and $W_0(x)\to\infty$ as $\omega(x)\to\infty$, such that $W(\cdot)=\pi\circ W_0(\cdot)$. We have the following Lyapunov-like lemma for \russonew{0-UGAS} hybrid systems \liu{adapted from \cite[Proposition 2]{NN-AK-RG:17}}.
\begin{lem}\label{lem:semiproper_V1}
   \liu{Let $\cA\subset\cX$ be a nonempty compact set.} If a hybrid system $\cH$ is \russonew{0-UGAS} with respect to $\cA$, then there exist a smooth semiproper function $V_1:\cX\to\Rp$, a function $\lambda\in\cK$ and a function $\rho\in\mathcal{PD}$ such that
    	\begin{align}
	    \max_{f\in\cF(x,u)}\langle\nabla V_1(x),f \rangle&\leq -\rho(\omega(x)) + \lambda(|u|) \nonumber \\ &  \forall (x,u)\in\cC,  \label{semiproper_c}\\
	 	\max_{g \in \cG(x,u)} V_1(g)-V_1(x)&\leq -\rho(\omega(x)) + \lambda(|u|)\nonumber \\ & \forall (x,u)\in\cD. \label{semiproper_d}
	\end{align}
\end{lem}

\subsection{Necessary conditions for UBEBS}\label{subsec:UBEBS}
In the previous subsection we have introduced two necessary conditions for \russonew{0-UGAS} hybrid system. In this subsection some necessary conditions for UBEBS will be presented, provided that the hybrid system is \russonew{0-UGAS}. The first result is similar to \cite[Lemma 2.1]{DA-ES-YW:00b}, characterizing UBEBS of hybrid systems using different estimates.
	\begin{lem}\label{lem:UBEBS_equivalent}
		\liu{Let $\cA\subset\cX$ be a nonempty compact set.} Suppose that a hybrid system $\cH$ is \russonew{0-UGAS} with respect to $\cA$. Then the following properties are equivalent:
		\begin{enumerate}[label=\alph*)]
			\item The system satisfies along all trajectories an estimate of the following type, for suitable functions $\alpha_1,\alpha_2,\alpha_3\in\cK_\infty$ and $\hat\gamma\in\cK$:
			\begin{equation}\label{eqn:UBEBS_equivalent_1}
				\!\!\!\!\alpha_1\big(\omega(x(t,j;x_0,u))\big)\!\leq\! \alpha_2(\omega(x_0))\!+\! \alpha_3\big(\Vert u\Vert^{\hat\gamma}_{(t,j)}\big).
			\end{equation}
			\item The system satisfies along all trajectories an UBEBS-like estimate with $c = 0$:
			\begin{equation}\label{eqn:UBEBS_equivalent_2}
				\alpha\big(\omega(x(t,j;x_0,u))\big)\leq \kappa(\omega(x_0))+ \Vert u\Vert^{\gamma}_{(t,j)}.
			\end{equation} 
		\item The system is UBEBS.
		\end{enumerate}
	\end{lem}

\begin{pf}
    To show the implication $a) \Rightarrow b)$, we start by assuming that property \eqref{eqn:UBEBS_equivalent_1} holds along all trajectories. Then \eqref{eqn:UBEBS_equivalent_2} holds with $\alpha(r):=\alpha_3^{-1}(\frac{1}{2}\alpha_1(r))$ and $\kappa(r):=\alpha_3^{-1}\circ\alpha_2(r)$. Meanwhile, the implication $b) \Rightarrow c)$ is trivial, by simply taking $c=0$. Thus, all that we need to prove is the implication $c) \Rightarrow a)$.
	   
Recall that \russonew{0-UGAS} implies local iISS by Lemma \ref{lem:0GAS_to_LiISS}. As a consequence, there exist functions $\beta\in\KLL$, $\chi\in\cK_{\infty},\gamma^a\in\cK$ and $l>0$ such that \eqref{eqn:iISS} holds
for all $x_0\in\B(\cA,l),u \in \cL_{\gamma^a}(l)$ and $(t,j)\in \dom x$. Define $\alpha_1^a(r):=\chi^{-1}(\frac{r}{2})$ and $\alpha_2^a(r):=\chi^{-1}\circ\beta(r,0,0)$, being both class $\cK_{\infty}$ functions. We have
\begin{equation}\label{eq:lemmaUBEBS_liISS}
    \alpha_1^a(\omega(x(t,j)))\leq \alpha_2^a(\omega(x_0))+\Vert u\Vert_{(t,j)}^{\gamma^a},
\end{equation}
for all $x_0\in\B(\cA,l), u \in \cL_{\gamma^a}(l), (t,j)\in \dom x$. Furthermore, the UBEBS property guarantees the existence of functions $\alpha_1^b,\alpha_2^b\in\cK_{\infty}, \gamma^b \in \cK$ and $c>0$ such that
\begin{equation}\label{eq:lemmaUBEBS_UBEBS}
    \alpha_1^b(\omega(x(t,j)))\leq \alpha_2^b(\omega(x_0))+\Vert u\Vert^{\gamma^b}_{(t,j)}+c,
\end{equation}
for all $(t,j)\in \dom x$. Now define $\alpha_1(s)=\min\lbrace\alpha_1^a(s),\alpha_1^b(s)\rbrace$, $\hat\alpha_2(s)=\max\lbrace\alpha_2^a(s),\alpha_2^b(s)\rbrace$ and $\hat\gamma(s)=\max\lbrace\gamma^a(s),\gamma^b(s)\rbrace$. Inequalities \eqref{eq:lemmaUBEBS_liISS} and \eqref{eq:lemmaUBEBS_UBEBS} respectively imply that
\begin{equation}\label{eq:0GAS+UBEBS=iISS_liISS}
    \alpha_1(\omega(x(t,j))) \leq \hat\alpha_2(\omega(x_0))+\Vert u\Vert^{\hat\gamma}_{(t,j)},
\end{equation}
for all $x_0\in\B(\cA,l), u \in \cL_{\hat\gamma}(l), (t,j)\in \dom x$ and 
\begin{equation}\label{eq:0GAS+UBEBS=iISS_UBEBS}
    \alpha_1(\omega(x(t,j))) \leq \hat\alpha_2(\omega(x_0))+\Vert u\Vert^{\hat\gamma}_{(t,j)}+c,
\end{equation}
for all $x\in\cX,u\in\cL_{\hat\gamma}^e,(t,j)\in \dom x$.
Further define the following functions,
\begin{align*}
    \alpha_2(s) &:= \begin{cases}
    \max\lbrace\hat\alpha_2(s),\frac{\hat\alpha_2(l)+c}{l}s\rbrace \quad &\text{if} \enskip s\leq l, \\
    \hat\alpha_2(s)+c\quad &\text{if} \enskip s> l, \\
    \end{cases}\\
    \alpha_3(s) &:= \begin{cases}
    (1+\frac{c}{l})s \quad &\text{if} \enskip s\leq l, \\
    s+c\quad &\text{if} \enskip s> l. \\
    \end{cases}
\end{align*}
By definition, both $\alpha_2,\alpha_3\in\cK_\infty$. Then, the following observations hold:
\begin{enumerate}
    \item If both $x_0\in\B(\cA,l), u \in \cL_{\hat\gamma}(l)$, then $\alpha_2(\omega(x_0))\geq\hat\alpha_2(\omega(x_0))$, $\alpha_3(\Vert u\Vert_{(t,j)}^{\hat\gamma})\geq \Vert u\Vert_{(t,j)}^{\hat\gamma}$. Hence \eqref{eqn:UBEBS_equivalent_1} follows from \eqref{eq:0GAS+UBEBS=iISS_liISS}.
    \item If $x_0\not\in\B(\cA,l), u \in \cL_{\hat\gamma}(l)$, then $\alpha_2(\omega(x_0))=\hat\alpha_2(\omega(x_0))+c$, $\alpha_3(\Vert u\Vert_{(t,j)}^{\hat\gamma})\geq \Vert u\Vert_{(t,j)}^{\hat\gamma}$. Hence \eqref{eqn:UBEBS_equivalent_1} follows from \eqref{eq:0GAS+UBEBS=iISS_UBEBS}.
    \item If $x_0\in\B(\cA,l), u \not\in \cL_{\hat\gamma}(l)$, then $\alpha_2(\omega(x_0))\geq\hat\alpha_2(\omega(x_0))$, $\alpha_3(\Vert u\Vert_{(t,j)}^{\hat\gamma})= \Vert u\Vert_{(t,j)}^{\hat\gamma}+c$. Hence \eqref{eqn:UBEBS_equivalent_1} again follows from \eqref{eq:0GAS+UBEBS=iISS_UBEBS}.
    \item If $x_0\not\in\B(\cA,l), u \not\in \cL_{\hat\gamma}(l)$, then $\alpha_2(\omega(x_0))=\hat\alpha_2(\omega(x_0))+c$, $\alpha_3(\Vert u\Vert_{(t,j)}^{\hat\gamma})=\Vert u\Vert_{(t,j)}^{\hat\gamma}+c$. Hence it follows from \eqref{eq:0GAS+UBEBS=iISS_UBEBS} that
    $\alpha_1(|x(t,j)|)\leq \hat\alpha_2(\omega(x_0))+\Vert u\Vert^{\hat\gamma}_{(t,j)}+2c =\alpha_2(\omega(x_0))+\alpha_3(\Vert u\Vert_{(t,j)}^{\hat\gamma})$, 
    so again we conclude \eqref{eqn:UBEBS_equivalent_1}.
\end{enumerate}
Given the above observations, the estimate \eqref{eqn:UBEBS_equivalent_1} holds for all $x_0\in\cX,u\in\cL_{\hat\gamma}^e$ and all $(t,j)\in\dom x$. This completes the implication $c) \Rightarrow a)$.\hfill$\blacksquare$
\end{pf}
	
From Lemma~\ref{lem:UBEBS_equivalent}, we see that if the system $\cH$ is UBEBS and \russonew{0-UGAS}, then an estimate of the form \eqref{eqn:UBEBS_equivalent_2} holds along its solutions. Such estimate also implies Lyapunov-like property for the hybrid system, which is stated by the next lemma.

	\begin{lem}\label{lem:UBEBS_Lyap}
		\liu{Let $\cA\subset\cX$ be a nonempty compact set.} Suppose that the solutions of the hybrid system $\cH$ satisfy an estimate as in \eqref{eqn:UBEBS_equivalent_2}. Then, there exist functions $\alpha_1',\alpha_2'\in\cK_\infty$ and a function $V_2 \colon \cX \to \R_{\geq 0}$ such that
		\begin{equation}\label{eq:UBEBS_Lyap_1} 
			\alpha_1'(\omega(x)) \leq V_2(x) \leq \alpha_2'(\omega(x))\quad\forall x\in \cX
		\end{equation}
		and for all $x_0\in\cX,u \in \cL_{\gamma}^e$, each solution pair $(x,u)\in\mathfrak s_u(x_0)$ satisfies
		\begin{equation}\label{eq:UBEBS_Lyap_2}
		    V_2(x(t,j))-V_2(x_0) \leq \Vert u \Vert_{(t,j)}^\gamma
		\end{equation}
		for all $(t,j)\in\dom x$.
	\end{lem}
	\begin{pf}
		Define $V_2$ as the following function:
  		\begin{equation}\label{def:V_2}
			V_2(x_0) \coloneqq \sup_{\scriptsize x,u,t,j} \left\lbrace\alpha(\omega(x(t,j;x_0,u)))
			-\Vert u \Vert_{(t,j)}^\gamma \right\rbrace,  
		\end{equation}
    where the decision variables are taken such that $u \in \cL_{\gamma}^e$, \liu{$(x,u)\in\mathfrak s_u(x_0)$}, $(t,j)\in\dom x$.  Consider $u=0, (t,j)=(0,0)$. We conclude that $V_2(x_0)\geq\alpha(\omega(x_0))$. Meanwhile, it follows from \eqref{eqn:UBEBS_equivalent_2} that $V_2(x_0)\leq\kappa (\omega(x_0))$. Thus $V_2$ is finite-valued; in particular, \eqref{eq:UBEBS_Lyap_1} is satisfied with $\alpha_1':=\alpha$ and $\alpha_2':=\kappa$.
		
		We next show \eqref{eq:UBEBS_Lyap_2}. 
		\liu{Pick arbitrary $(x,u)\in\mathfrak s_u(x_0)$ and $(t,j)\in\dom x$. Denote $x^*:=x(t,j;x_0,u)$. It follows from the definition of $V_2$ in \eqref{def:V_2} that for any $\epsilon>0$, there exist $(\tilde x,\tilde u)\in\mathfrak s_{\tilde u}(x^*)$ and $(\tau,k)\in\dom \tilde x$ such that
		\begin{equation}\label{a_step_for_V_2}
		V_2(x^*)\leq \alpha(\omega(\tilde x(\tau,k;x^*,\tilde u)))-\Vert\tilde u\Vert_{(\tau,k)}^\gamma+\epsilon.		    
		\end{equation}
		Now define a hybrid input $u\#\tilde u$ such that $\dom(u\#\tilde u):=\dom u\cup\{(\tau,k)\in\R\times\N:(\tau-t,k-j)\in\dom\tilde u\}$, and
		\begin{equation*}
		    (u\#\tilde u)(\tau,k):=\begin{cases}
		    u(\tau,k)&\mbox{ if } (\tau,k)\leq (t,j),\\
		    \tilde u(\tau-t,k-j)&\mbox{ otherwise}.
		    \end{cases}
		 \end{equation*}
		 In other words, $u\#\tilde u$ is the concatenation of $u$ and $\tilde u$. It is not difficult to see that $\Vert u\#\tilde u\Vert_{(t+\tau,j+k)}^\gamma=\Vert u\Vert_{(t,j)}^\gamma+\Vert \tilde u\Vert_{(\tau,k)}^\gamma$. Meanwhile, since $\tilde x(\tau,k;x^*,\tilde u)=x(t+\tau,k+j;x_0,u\#\tilde u)$, It follows from \eqref{def:V_2} and \eqref{a_step_for_V_2} that
		 \begin{align*}
		     V_2(x^*)&\leq V_2(x_0)+\Vert u\#\tilde u\Vert_{(t+\tau,j+k)}^\gamma-\Vert \tilde u\Vert_{(\tau,k)}^\gamma+\epsilon\\
		     &= V_2(x_0)+\Vert u\Vert_{(t,j)}^\gamma+\epsilon.
		 \end{align*}
		 Because $\epsilon$ can be arbitrarily chosen, we must have $V_2(x^*)\leq V_2(x_0)+\Vert u\Vert_{(t,j)}^\gamma$, which proves this lemma.
		\hfill$\blacksquare$}
	\end{pf}

\section{\liu{A non-smooth Lyapunov characterization} for hybrid systems}\label{sec:non-smooth_V}
In Section~\ref{subsec:UBEBS}, we defined a function $V_2$ satisfying the inequality \eqref{eq:UBEBS_Lyap_2} when the hybrid system $\cH$ is both \russonew{0-UGAS} and UBEBS. This inequality does not necessarily guarantee that $V_2$ is decreasing along solutions when $u=0$. However, if $V_2$ is smooth, then we can further conclude from \eqref{eq:UBEBS_Lyap_2} that
\begin{align*}
    \max_{f\in \cF(x,u)}\langle \nabla V_2(x),f\rangle \leq \gamma(|u|)\quad\forall (x,u)\in\cC, \\
    \max_{g\in \cG(x,u)}V_2(g)-V_2(x)\leq \gamma(|u|)\quad\forall (x,u)\in\cD.
\end{align*}
\russo{Note that the above condition is reminiscent of the zero-output dissipativity property considered in \cite{DA-ES-YW:00}}. Recall the function $V_1$ from Lemma~\ref{lem:semiproper_V1} which has the properties~\eqref{semiproper_c}, \eqref{semiproper_d}. Thus, by defining $\bar V:=V_1+V_2$, we conclude that
\begin{align}
       \max_{f\in \cF(x,u)}\langle \nabla \bar V(x),f\rangle &\leq -\rho(\omega(x))+\sigma(|u|)\nonumber\\
       &\forall (x,u)\in\cC, \label{smooth_V_c}\\
    \max_{g\in \cG(x,u)}\bar V(g)-V_2(x)&\leq -\rho(\omega(x))+\sigma(|u|)\nonumber\\
    &\forall (x,u)\in\cD,\label{smooth_V_d}
\end{align}
where $\sigma:=\lambda+\gamma\in\cK$. The inequalities \eqref{smooth_V_c}, \eqref{smooth_V_d} are necessary conditions for $\bar V$ to be an iISS Lyapunov function and show iISS for the system $\cH$ \cite[Theorem 1]{NN-AK-RG:17}. Nevertheless, this approach is infeasible in our work, since Lemma~\ref{lem:UBEBS_Lyap} does not guarantee smoothness of $V_2$. We therefore aim to prove iISS using an alternative approach, which relies on a non-smooth version of iISS Lyapunov characterization for hybrid systems.


\begin{lem}\label{thm:non_smooth_iISS_V}
    \liu{Let $\cA\subset\cX$ be a nonempty compact set.} Consider a hybrid system $\cH$ and suppose that the Standing Assumptions hold. The system $\cH$ is iISS if and only if there exist a function $\bar V(x):\cX\to\Rp$, functions $\bar\alpha_1,\bar\alpha_2\in\cK_{\infty}, \bar\gamma\in\cK$ and $\rho\in\mathcal{PD}$ such that
    	    \begin{align}
        \bar\alpha_1(\omega(x))&\leq \bar V(x)\leq \bar\alpha_2(\omega(x))\quad\forall x\in\cX,\label{special_sandwich}\\
        \bar V(x(t,j;x_0,u))&\leq \bar V(x_0)\!-\!\int_0^t\!\rho( \omega(x(\tau,i(\tau);x_0,u)))\,\mathrm{d}\tau \nonumber\\
        &-\hspace{-1cm} \sum_{\tiny\begin{array}{c}
        		(t',j')\in\Gamma(x), \\
        		(0,0)\leq(t',j')\leq(t,j)
        \end{array}}\hspace{-1cm}\rho(\omega(x(t',j';x_0,u)))+\Vert u\Vert_{(t,j)}^{\bar\gamma}\nonumber\\
    &\forall x_0\in\cX,u\in\cL_{\bar\gamma}^e,(t,j)\in \dom x.\label{special_V}
    \end{align}
\end{lem}
\begin{rem}
	The novelty of Lemma~\ref{thm:non_smooth_iISS_V} is that iISS does not require $\bar V$ to be smooth (and hence $\bar V$ is not necessarily a smooth iISS Lyapunov function in the sense of \cite[Definition 2]{NN-AK-RG:17}). One could of course use similar techniques as in \cite{DA-ES-YW:00b} to majorize and smooth $\bar V$ so that the inequalities \eqref{smooth_V_c}, \eqref{smooth_V_d} still hold; however, in the framework of hybrid systems, such an approach becomes significantly more complicated because of the continuous flow/discrete jump mixed dynamics.

\end{rem}
Prior to proving the non-smooth Lyapunov characterization, some required technical results need to be introduced.

\subsection{Technical tools for proving the non-smooth iISS Lyapunov characterization}
The first lemma shows that the estimates in terms of $\KLL$ functions are essentially not different to the estimates in terms of $\KL$ functions. 

\begin{lem}\label{lem:equivalent_KL_and_KLL}
	For any $\tilde\beta\in\KL$, there exists $\beta\in\KLL$ such that $\tilde\beta(s,t+j)\leq\beta(s,t,j)$ for any $s,t,j\geq 0$. Similarly for any $\beta\in\KLL$, there exists $\tilde\beta\in\KL$ such that $\beta(s,t,j)\leq\tilde\beta(s,t+j)$ for any $s,t,j\geq 0$.
\end{lem}
\ifthenelse{\equal{\complete}{T}}{
 \begin{pf}
 	To show the first part, let $\beta(s,t,j):=\tilde \beta(s,t+j)$. Then clearly we have $\beta\in\KLL$ and $\tilde\beta(s,t+j)\leq\beta(s,t,j)$ (in fact equality holds) for any $s,t,j\geq 0$. To show the second part, let $s,t,j\geq 0$ be arbitrary and denote $z=t+j$. Define $\tilde\beta(s,z):=\beta(s, \frac{z}{2},0)+\beta(s,0,\frac{z}{2})$. By definition $\tilde\beta\in\KL$. When $t\geq \frac{z}{2}$, we have $\beta(s,t,j)\leq \beta(s,\frac{z}{2},0)\leq\tilde\beta(s,z)$. On the other hand, if $t<\frac{z}{2}$, one must have $j>\frac{z}{2}$ so 	$\beta(s,t,j)\leq \beta(s,0,\frac{z}{2})\leq\tilde\beta(s,z)$. In both cases we conclude $\beta(s,t,j)\leq\tilde\beta(s,t+j)$.\hfill$\blacksquare$
 \end{pf}}{
Proof of Lemma \ref{lem:equivalent_KL_and_KLL} can be found in \cite{SL-AR:22-arxiv}.}
The next lemma provides an equivalent characterization of iISS from the analysis point of view.

\begin{lem}\label{lem:special_char_hybrid_iISS}
	\liu{Let $\cA\subset\cX$ be a nonempty compact set.} Consider a hybrid system $\cH$ and suppose that the Standing Assumptions hold. The system $\cH$ is iISS if and only if there exists $\gamma\in\cK$ such that the following \russonewnew{three} properties hold:
	\begin{enumerate}
		\item[1.] For each $\epsilon>0$, there exists a $\delta>0$ such that
		\begin{equation}\label{eqn:property_1}
			\omega(x(t,j;x_0,u))\leq\epsilon
		\end{equation}
		for all $x_0\in\B(\cA,\delta),u\in\cL_\gamma(\delta)$ and $(t,j)\in\dom x$. 
		\item[2.] There exists $\chi\in\cK_{\infty}$ such that for any $r,\epsilon>0$, there exists $Z>0$ such that for each $u\in\cL_\gamma^e$,
		\begin{equation}\label{eqn:property_2}
			\omega(x(t,j;x_0,u))\leq\epsilon+\chi(\Vert u\Vert_{(t,j)}^\gamma)
		\end{equation}
		for all $x_0\!\in\B(\cA,r)$ and $(t,j)\in\dom x$ with $t+j\geq\! Z$.
        \item[3.] \russonewnew{System $\cH$ is UBEBS}.
	\end{enumerate}
\end{lem}

\begin{rem}
	When Properties 2 \russonewnew{and 3} in Lemma~\ref{lem:special_char_hybrid_iISS} holds, Property 1 can be replaced by the following:
	\textit{\begin{enumerate}
		\item[1'.] There exists $\delta,\tilde\chi\in\cK_{\infty}$ such that
		\begin{equation}\label{eqn:property_1prime}
			\omega(x(t,j;x_0,u))\leq\epsilon+\tilde\chi(\Vert u\Vert_{(t,j)}^\gamma)
		\end{equation}
		for all $x_0\in\B(\cA,\delta(\epsilon))$ and $(t,j)\in\dom x$.
	\end{enumerate}}
	This can be proven by a similar approach in \cite[Remark 2.8]{ES-YW:95_2}. \russonewnew{To this end, we need to show that for any positive scalars $Z$, $r$ and $s$, there exists $L>0$ such that $\omega(x(t,j;x_0,u))\leq L$ for all $t+j\leq Z$, $x_0\in \B(\cA,r)$, $u \in \cL_\gamma(s)$. This property (which is similar to the bounded reachable set (BRS) property in \cite[Definition 4]{AM-FW:18}) is directly concluded from \eqref{eqn:UBEBS} under the assumption that the hybrid system is \russonewnew{UBEBS}.}
\end{rem}

We postpone its proof after a discussion on the next lemma, which is essentially a refinement of Property 2 in Lemma~\ref{lem:special_char_hybrid_iISS} by providing more details regarding the relationships between $Z$ and $r,\epsilon$.
\begin{lem}\label{lem:Refinement_of_Z}
	When both Property 1 and 2 in Lemma~\ref{lem:special_char_hybrid_iISS} hold, Property 2 is equivalent to the following: there exists $\chi\in\cK_{\infty}$ and a family of mappings $\{Z_r\}_{r>0}$ with the following properties:
	\begin{itemize}
		\item for each fixed $r>0$, $Z_r:\Rsp\to\Rsp$ is subjective, continuous and strictly decreasing. In particular, $\lim_{s\to\infty}Z_r(s)=0$.
		\item for each fixed $\epsilon>0$, $Z_r(\epsilon)$ is strictly increasing as $r$ increases and $\lim_{r\to\infty} Z_r(\epsilon)=\infty$.
		\item for each $u\in\cL_\gamma^e$, the estimation \eqref{eqn:property_2} holds for all $x_0\in\B(\cA,r)$ and $(t,j)\in\dom x$ with $t+j\geq Z_r(\epsilon)$.
	\end{itemize}
\end{lem}

\ifthenelse{\equal{\complete}{T}}{
\begin{pf}
	Sufficiency is clear. The necessity part is very similar to the proof of \cite[Lemma 3.1]{YL-EDS-YW:96} so only a sketch is given. Let $\chi\in\cK_{\infty}$ be given as in \eqref{eqn:property_2}. For each $u\in\cL_\gamma^e$ and $r,\epsilon>0$, let 
	\begin{multline*}
		\tilde Z_{r,\epsilon}:=\inf\{Z:\eqref{eqn:property_2} \mbox{ holds for all } x_0\in\B(\cA,r),\\
		(t,j)\in\dom x,t+j\geq Z\}.
	\end{multline*}
	We have $\tilde Z_{r,\epsilon}<\infty$ for any $r,\epsilon>0$. Further, $\tilde Z_{r_1,\epsilon}\leq \tilde Z_{r_2,\epsilon}$ if $r_1\leq r_2$, and $\tilde Z_{r,\epsilon_1}\geq \tilde Z_{r,\epsilon_2}$ if $\epsilon_1\leq \epsilon_2$. Also, Property 1' implies that, for every fixed $r$, $\tilde Z_{r,\epsilon}\to 0$ as $s\to\infty$. Now for each $r,s>0$, let $\bar Z_r(s)=\frac{2}{s}\int_{\frac{s}{2}}^s\tilde Z_{r,\tau}\,\mathrm{d}\tau$. Then, for each fixed $r>0$, $\bar Z_r$ is a continuous function, and $\bar Z_r(s)\geq \tilde Z_{r,s}$. Finally, for each $r,s>0$, let $Z_r(s):=\frac{r}{s}+\sup_{\tau\geq s}\bar Z_r(\tau)$. One can easily check that the family $\{Z_r\}_{r\geq 0}$ satisfies all the conditions in the lemma.
\end{pf}
}{\liu{The proof is similar to the one for \cite[Lemma 3.1]{YL-EDS-YW:96} and hence omitted here.}}

\begin{pf*}{Proof of Lemma~\ref{lem:special_char_hybrid_iISS}.}
	The necessity part is clear. Assume now that the two properties in Lemma~\ref{lem:special_char_hybrid_iISS} hold. Let $\delta\in\cK_{\infty}$ be as in Property 1', and without loss of generality, we assume that $\tilde\chi$ in Property 1' and $\chi$ in Property 2 are the same function, denoted by $\chi$ and majorized to be a class $\cK_{\infty}$ function. On one hand, let $\varphi=\delta^{-1}$. Then it holds that
	\begin{equation}\label{eqn:varphi_part}
		\omega(x(t,j;x_0,u))\leq\varphi(\omega(x_0))+\chi(\Vert u\Vert_{(t,j)}^\gamma)
	\end{equation}
	for all $(t,j)\in\dom x$. On the other hand, let $\{Z_r\}_{r\geq 0}$ be as in Lemma~\ref{lem:Refinement_of_Z}, and for each $r>0$, let \begin{equation*}
		\psi_r(s):=\begin{cases}
			Z_r^{-1}(s)&\mbox{ if }s>0,\\
			\infty&\mbox{ if }s=0.
		\end{cases}
	\end{equation*}
	Note then that $\psi_r$ is continuous on $(0,\infty)$ and $\lim_{s\to 0}\psi_r(s)=\infty$ for each $r>0$. Since \eqref{eqn:property_2} holds when $t+j\geq Z_r(\epsilon)$, $\omega(x_0)\leq r$, and $s=Z_r(\psi_r(s))$ for any $s>0$, it follows from the above, applied in the particular case for $t+j=Z_r(\epsilon)$ that for any $(t,j)>(0,0)$,
	\begin{equation}\label{eqn:psi_part}
		\omega(x(t,j;x_0,u))\leq\psi_r(t+j)+\chi(\Vert u\Vert_{(t,j)}^\gamma)
	\end{equation}
	for any $u\in\cL_\gamma^e$ and any $\omega(x_0)\leq r$. This formula also holds for $(t,j)=(0,0)$ by the definition that $\psi_r(0)=\infty$. 
	Now for any $s,t\geq 0$, let $\bar\psi(s,t):=\min\{\inf_{r\geq s}\psi_r(t),\varphi(s)\}$. Then by \eqref{eqn:varphi_part} and \eqref{eqn:psi_part}, one has
	\begin{equation}\label{eqn:combining_varphi_psi}
		\omega(x(t,j;x_0,u))\leq\bar\psi(\omega(x_0),t+j)+\chi(\Vert u\Vert_{(t,j)}^\gamma).
	\end{equation}
	Pick any function $\tilde\psi:\Rp\times\Rp\to\Rp$ with the following properties:
	\begin{enumerate}
		\item For any fixed $z\geq 0$, $\tilde\psi(\cdot,z)$ is continuous and strictly increasing.
		\item For any fixed $s\!\geq\! 0$, $\tilde\psi(s,z)$ decreases to $0$ as $z\!\to\!\infty$.
		\item $\tilde\psi(s,z)\geq\bar\psi(s,z)$ for all $s,z\geq 0$.
	\end{enumerate}
	Such majorized function $\tilde\phi$ for $\bar\phi$ always exists, by following the similar proof of \cite[Proposition 2.5]{YL-EDS-YW:96}. Define $\tilde\beta(s,z):=\sqrt{\varphi(s)\tilde\psi(s,z)}$. Then $\tilde\beta\in\KL$, and moreover,
	\begin{equation}\label{eqn:property_tilde_beta}
		\tilde\beta(s,z)\geq \min\{\varphi(s),\tilde\psi(s,z)\}
	\end{equation}
	for all $s,z\geq 0$. Combining \eqref{eqn:combining_varphi_psi} and \eqref{eqn:property_tilde_beta}, one concludes that $\omega(x(t,j;x_0,u))\leq\tilde\beta(\omega(x_0),t+j)+\chi(\Vert u\Vert_{(t,j)}^\gamma)$ for all $(t,j)\in\dom x, x_0\in\cX$ and all $u\in\cL_\gamma^e$. Finally, it follows from Lemma~\ref{lem:equivalent_KL_and_KLL} that there exists $\beta\in\KLL$ such that $\tilde\beta(s,t+j)\leq\beta(s,t,j)$ and we conclude \eqref{eqn:iISS}. Hence the system is iISS.\hfill$\blacksquare$
\end{pf*}

\subsection{Proof for the non-smooth Lyapunov characterization}
We are now ready to prove Lemma~\ref{thm:non_smooth_iISS_V}. 

\begin{pf*}{Proof of Lemma~\ref{thm:non_smooth_iISS_V}.}
    To show the necessity part, we start with assuming that the system $\cH$ is iISS. It follows from \cite[Theorem 1]{NN-AK-RG:17} that there exist a smooth $\bar V:\cX\to\Rp$, $\bar\alpha_1,\bar\alpha_2\in\cK_\infty$, $\sigma\in\cK$ and $\rho\in\PD$ such that \eqref{smooth_V_c},\eqref{smooth_V_d} and \eqref{special_sandwich} hold. Along an arbitrary hybrid solution of the system $\cH$, integrate \eqref{smooth_V_c} for $(t,j)\in\dom x\backslash\Gamma(x)$ and sum up \eqref{smooth_V_d} for $(t,j)\in\Gamma(x)$, we conclude \eqref{special_V} with $\bar\gamma:=\sigma$.
    
    To show the sufficiency part, we start with assuming that there exist $\bar V:\cX\to\Rp$, $\bar\alpha_1,\bar\alpha_2\in\cK_\infty$, $\sigma\in\cK$ and $\rho\in\PD$ such that \eqref{special_sandwich} and \eqref{special_V} hold. Let $\gamma:=\bar\gamma$. It follows from \eqref{special_sandwich} and \eqref{special_V} that for any $(t,j)\in\dom x$,
    \begin{align}
        \omega(x(t,j))&\leq \bar\alpha_1^{-1}(\bar V(x(t,j)))\nonumber \\
        &\leq \bar\alpha_1^{-1}\big(\bar V(x_0)+\Vert u\Vert_{(t,j)}^{\gamma}\big)\nonumber\\
        &\leq\bar\alpha_1^{-1}\big(2\bar V(x_0)\big)+\bar\alpha_1^{-1}\big(2\Vert u\Vert_{(t,j)}^{\gamma}\big)\nonumber\\
        &\leq \bar\alpha_1^{-1}\big(2\bar\alpha_2(\omega(x_0))\big)+\bar\alpha_1^{-1}\big(2\Vert u\Vert_{(t,j)}^{\gamma}\big)\label{estimate_on_x(t)}.
    \end{align}
    To show that Property 1 in Lemma~\ref{lem:special_char_hybrid_iISS} holds for $\cH$, let $\delta=\min\{\bar\alpha_2^{-1}(\frac{1}{2}\bar\alpha_1(\frac{\epsilon}{2})),\frac{1}{2}\bar\alpha_1(\frac{\epsilon}{2})\}$. It directly follows from \eqref{estimate_on_x(t)} that when $x_0\in\B(\cA,\delta)$ and $u\in\cL_\gamma(\delta)$, \eqref{eqn:property_1} holds for all $(t,j)\in\dom x$. Note that this property can also be directly concluded if $\cH$ is assumed to be locally iISS. To show that Property 2 in Lemma~\ref{lem:special_char_hybrid_iISS} holds for $\cH$, let $\chi(s):=\bar\alpha_1^{-1}(2s)$. If $\epsilon\geq\bar\alpha_1^{-1}(2\bar\alpha_2(r))$, it again follows from \eqref{estimate_on_x(t)} that when $x_0\in\B(\cA,r)$, one has
    \begin{equation*}
        \omega(x(t,j)) \!\leq\! \bar\alpha_1^{-1}\big(2\bar\alpha_2(r)\big)\!+\!\bar\alpha_1^{-1}\big(2\Vert u\Vert_{(t,j)}^{\gamma}\big) \!\leq\! \epsilon \!+\!\chi(\Vert u\Vert_{(t,j)}^{\gamma})
    \end{equation*}
    for all $(t,j)\in \dom x$. Hence in this case the inequality \eqref{eqn:property_2} holds with $t+j\geq Z:=0$. Otherwise, if $\epsilon<\alpha_1^{-1}(2\alpha_2(r))$, then
    \begin{equation*}
        a\!\coloneqq\!\bar\alpha_2^{-1}\big(\frac{1}{2}\bar\alpha_1(\epsilon)\big)\!\leq\! \bar\alpha_1^{-1}\big(\frac{1}{2}\bar\alpha_1(\epsilon)\big)\!\leq\! \epsilon\!<\! \bar\alpha_1^{-1}(2\bar\alpha_2(r))\!=:\!b.
    \end{equation*}
    Define $\underline\rho:=\min_{s\in[a,b]}\rho(s)>0$ and $Z:= \frac{2\bar\alpha_2(r)}{\underline\rho} +1$. We proceed the proof by discussing two cases: comparing $\Vert u\Vert_{(t,j)}^\gamma$ and $\bar\alpha_2(r)$. If $\Vert u\Vert_{(t,j)}^\gamma<\bar\alpha_2(r)$, then suppose $\bar V(x(t,j))\geq \frac{1}{2}\bar\alpha_1(\epsilon)$ for all $(t,j)\in\dom x$ with $t+j\leq Z$, in which case we have
    \begin{equation}\label{bound_a}
        \omega(x(t,j))\geq\bar\alpha_2^{-1}(\bar V(x(t,j)))\geq \bar\alpha_2^{-1}\big(\frac{1}{2}\bar\alpha_1(\epsilon)\big)=a.
    \end{equation} 
    Moreover, properties \eqref{special_sandwich} and \eqref{special_V} further imply that
    \begin{align*}
        \bar V(x(t,j))&\leq \bar V(x_0)+\Vert u\Vert_{(t,j)}^\gamma\\
        &\leq \bar\alpha_2(\omega(x_0))+\Vert u\Vert_{(t,j)}^\gamma < 2\bar\alpha_2(r)
    \end{align*}
    for all $(t,j)\in\dom x$ with $t+j\leq Z$. Hence
    \begin{equation}\label{bound_b}
        \omega(x(t,j))\leq\bar\alpha_1^{-1}(\bar V(x(t,j))\leq \bar\alpha_1^{-1}(2\bar\alpha_2(r))=b.
    \end{equation}
    Therefore, \eqref{bound_a} and \eqref{bound_b} imply $\omega(x(t,j))\in[a,b]$ and hence $\rho(\omega(x(t,j)))\geq\underline\rho$ for all $(t,j)\in\dom x$ with $t+j\leq Z$. 
    Define 
    \begin{equation}\label{def:S}
        S(r):=\{(t,j)\in\dom x:t+j\in[r-1,r]\},
    \end{equation}
    which is non-empty for any $r\in[1,\length(\dom x)+1)$. Thus the estimate \eqref{special_V} with $(t,j)=(T,J)\in S(Z)$ can be approximated slightly tighter:
    \begin{align*}
        \bar V(x(T,J))&\leq \bar\alpha_2(\omega(x_0))-\underline\rho T -\underline\rho J+\Vert u\Vert_{(T,J)}^\gamma \\
        &< 2\bar\alpha_2(r)-\underline\rho (Z-1)= 0.
    \end{align*}
    However since $T+J\leq Z$, we should have $\bar  V(x(T,J))\geq \frac{1}{2}\bar\alpha_1(\epsilon)$ by the hypothesis, which is a contradiction. Therefore we must have $\bar V(x(t^*,j^*))< \frac{1}{2}\bar\alpha_1(\epsilon)$ for some $(t^*,j^*)\in\dom x$ with $t^*+j^*\leq Z$. Since the system is time-invariant, \eqref{special_V} should also hold with initial hybrid time being $(t^*,j^*)$, that is, for any $(t,j)\in\dom x$ with $t+j\geq Z$,
\begin{align*}
    \bar V(x(t,j)) &\leq \bar V(x(t^*,j^*))-\int_{t^*}^t\rho(\omega(x(\tau, i(\tau))))\,\mathrm{d}\tau\\
    &- \hspace{-0.75cm} \!\!\sum_{\tiny\begin{array}{c}
					(t',j')\in\Gamma(x),  \\
					(t^*,j^*)\leq(t',j')\leq(t,j)
			\end{array}} \hspace{-1cm} \!\! \rho(\omega(x(t',j')))\!+\!\int_{t^*}^t\!\!\!\gamma(|u(\tau,i(\tau))|)\,\mathrm{d}\tau \\
    &+ \hspace{-0.75cm} \!\!\sum_{\tiny\begin{array}{c}
					(t',j')\in\Gamma(u),  \\
					(t^*,j^*)\leq(t',j')\leq(t,j)
			\end{array}} \hspace{-1cm} \gamma\big(|u(t',j')|\big) \!\leq\! \bar V(x(t^*,j^*))\!+\!\Vert u\Vert_{(t,j)}^\gamma.
\end{align*}
Hence similar to the derivation for \eqref{estimate_on_x(t)}, one concludes that
\begin{align*}
    \omega(x(t,j))&\leq \bar\alpha_1^{-1}\big(2\bar V(x(t^*,j^*))\big)+\bar\alpha_1^{-1}\big(2\Vert u\Vert_{(t,j)}^\gamma\big) \\
    &\leq\epsilon+\chi(\Vert u\Vert_{(t,j)}^\gamma).
\end{align*}
In the other case, $\Vert u\Vert_{(t,j)}^\gamma\geq\bar\alpha_2(r)\geq \bar\alpha_2(\omega(x_0))$ and we directly conclude that for any $(t,j)\in\dom x$ with $t+j\geq Z$,
\begin{align*}
               \omega(x(t,j))&\leq \bar\alpha_1^{-1}\big(\bar V(x(t,j))\big)\\
               &\leq \bar\alpha_1^{-1}\big(\bar V(x_0)+\Vert u\Vert_{(t,j)}^\gamma\big)\\
               &\leq \bar\alpha_1^{-1}\big(\bar\alpha_2(\omega(x_0))-\Vert u\Vert_{(t,j)}^\gamma+2\Vert u\Vert_{(t,j)}^\gamma\big)\\
               &\leq \bar\alpha_1^{-1}\big(2\Vert u\Vert_{(t,j)}^\gamma\big) \leq \epsilon+\chi(\Vert u\Vert_{(t,j)}^\gamma).
\end{align*}
\russonewnew{Finally, Property 3 in Lemma \ref{lem:special_char_hybrid_iISS} directly follows from \eqref{estimate_on_x(t)}.}
Therefore, \russonewnew{all the} properties in Lemma~\ref{lem:special_char_hybrid_iISS} are proven and hence $\cH$ is iISS.\hfill$\blacksquare$
\end{pf*}

\section{Proof for the main result}\label{sec:other}

In this section we provide the proof of Theorem~\ref{thm:main}. Notice that the implication $1) \Rightarrow 3)$ is trivial. The implication $2) \Rightarrow 1)$ and $3) \Rightarrow 2)$ will be done in two parts.



\begin{pf*}{Proof of $2) \Rightarrow 1)$.}
We start with assuming that $\cH$ is both \russonew{0-UGAS} and UBEBS. By Lemma~\ref{lem:0GAS_to_LiISS}, \russonew{0-UGAS} implies the existence of a smooth semiproper function $V_1:\cX\to\Rp$, a function $\lambda\in\cK$ and a function $\rho\in\mathcal{PD}$ such that \eqref{semiproper_c}, \eqref{semiproper_d} hold. Since $V_1$ is semiproper, there exists $\alpha_2''\in\cK_{\infty}$ (e.g., because $V_1:=\pi\circ V$ as in the proof of Lemma~\ref{lem:0GAS_to_LiISS}, we pick $\pi'\in\cK_{\infty}$ such that $\pi'(s)\geq \pi(s)$ for all $s\in\Rp$ and define $\alpha_2'':=\pi'\circ\alpha_2$, where recall $\alpha_2$ is an upper bound on $V$ as in \eqref{eq:lemmaUBEBS_1}) such that
\begin{equation}\label{alpha_2''}
    V_2(x)\leq \alpha_2''(\omega(x))\quad\forall x\in\cX.
\end{equation}
Along an arbitrary hybrid solution of the system $\cH$, integrate \eqref{semiproper_c} for $(t,j)\in\dom x\backslash\Gamma(x)$ and sum up \eqref{semiproper_d} for $(t,j)\in\Gamma(x)$, we conclude that
\begin{align}
        V_1(x(t,j))&\leq V_1(x_0)-\int_0^t\rho(\omega(x(\tau,i(\tau))))\,\mathrm{d}\tau \nonumber\\
        &-\hspace{-1cm} \sum_{\tiny\begin{array}{c}
        		(t',j')\in\Gamma(u), \\	(0,0)\leq(t',j')\leq(t,j)
        \end{array}}\hspace{-1cm}\rho(\omega(x(t',j')))+\Vert u\Vert_{(t,j)}^{\lambda}\label{step:V_1}
\end{align}
for all     $ x_0\in\cX,u\in\cL_{\gamma}^e,(t,j)\in \dom x$.
On the other hand, by Lemma~\ref{lem:UBEBS_equivalent} and Lemma~\ref{lem:UBEBS_Lyap} , \russonew{0-UGAS} and UBEBS imply the existence of functions $\alpha_1',\alpha_2'\in\cK_{\infty}$, a function $V_2:\cX\to\Rp$ such that \eqref{eq:UBEBS_Lyap_1}, \eqref{eq:UBEBS_Lyap_2} hold. Now define $\bar V:=V_1+V_2$.  It follows from \eqref{eq:UBEBS_Lyap_1} and \eqref{alpha_2''} that \eqref{special_sandwich} holds with $\bar\alpha_1:=\alpha_1',\bar\alpha_2:=\alpha_2'+\alpha_2''$. Moreover, it follows from \eqref{eq:UBEBS_Lyap_2} and  \eqref{step:V_1} that \eqref{special_V} holds with $\rho$ and $\bar\gamma:=\gamma+\lambda$. Hence by Lemma~\ref{thm:non_smooth_iISS_V}, the system~$\cH$ is iISS.\hfill$\blacksquare$
\end{pf*}

\begin{pf*}{Proof of $3) \Rightarrow 2)$ in Theorem~\ref{thm:main}.}
To prove this implication, we start with assuming $\cH$ to be $l$-locally iISS and $p$-practically iISS with $p>l$. Practical iISS means that estimate \eqref{eqn:practical_iISS}
holds along all hybrid solutions. By defining $\alpha(r):=\chi^{-1}(\frac{r}{2}),\kappa(r):=\chi^{-1}\circ\beta(r,0,0),c:=p$, we conclude \eqref{eqn:UBEBS}. In other words, the system $\cH$ is UBEBS. 

Next we show that $\cH$ is also \russonew{0-UGAS}. Consider the case when $u=0$. Inequality \eqref{eqn:practical_iISS} and Lemma~\ref{lem:equivalent_KL_and_KLL} imply the existence of $\tilde \beta_1\in\KL$ such that the estimate
\begin{equation}\label{case:practical}
    \omega(x(t,j;x_0,0))\leq \tilde\beta_1(\omega(x_0),t+j)+p
\end{equation}
holds for all unforced hybrid solutions with any $x_0\in\cX$.
Meanwhile, the inequality \eqref{eqn:iISS} for local iISS and Lemma~\ref{lem:equivalent_KL_and_KLL} implies the existence of $\tilde\beta_2\in\KL$ such that the estimate
\begin{equation}\label{case:local}
    \omega(x(t,j;x_0,0))\leq \tilde\beta_2(\omega(x_0),t+j)
\end{equation}
holds for all unforced hybrid solutions with any $x_0\in\B(\cA,l)$.
Note it is assumed that $l>p$. For each $s\in\Rp$, Define $T^*(s)\in\Rp$ such that $T^*(s)=0$ if $\tilde\beta_1(s,0)\leq l-p$ and $\tilde\beta_1(s,T^*(s))=l-p$ otherwise. Further define a function $\tilde\beta:\Rp\times\Rp\to\Rp$ by
\begin{equation*}
    \tilde\beta(s,t)\!:=\!\begin{cases}
    \frac{\tilde\beta_2(l,0)}{l}\left(\tilde\beta_1(s,t)+p\right)\!+\!\frac{1}{t}&\!\!\mbox{ if }t\!<\! T^*(s),\\
    \tilde\beta_2(l,\max\{t-T^*(s)-1,0\})\!+\!\frac{1}{t}&\!\!\mbox{ if }t\!\geq\! T^*(s).
    \end{cases}
\end{equation*}
 By this definition, $\tilde\beta(s,t)$ is continuous at $t=T^*(s)$. It is also not difficult to check that $\tilde\beta\in\KL$. We claim that 
\begin{equation}\label{case:combined}
    \omega(x(t,j;x_0,0))\leq \tilde\beta(\omega(x_0),t+j)
\end{equation}
for all $x_0\in\cX,(t,j)\in\dom x$; in other words, with the help of Lemma~\ref{lem:equivalent_KL_and_KLL}, we claim that the system~$\cH$ is \russonew{0-UGAS}. To prove the claim, we recall the definition of $S(r)$ from \eqref{def:S}. Let $x\in\mathfrak s_0(x_0)$ with $x_0\in\cX$. Let $(t^*,j^*)=\min\{ S(T^*(\omega(x_0))+1)\}$. For all $(t,j)\in\dom x$, if $t+j\leq T^*(\omega(x_0))$, then \eqref{case:combined} follows from \eqref{case:practical}. If  $t+j> T^*(\omega(x_0))$, then it follows from the definition of $(t^*,j^*)$ that $t^*+j^*\leq t+j$ and $t^*+j^*\leq T^*(s)+1$. It can also be concluded from \eqref{case:practical} that $\omega(x(t^*,j^*))\leq l$, hence by treating $x(t^*,j^*)$ as the initial state, the inequality \eqref{case:local} implies that
\begin{align*}
    \omega(x(t,j;x_0,0))&=\omega(x(t-t^*,j-j^*;x(t^*,j^*),0))\\
    &\leq \tilde\beta_2(\omega(x(t^*,j^*)),t-t^*+j-j^*)\\
    &\leq \tilde\beta_2(l,\max\{(t+j)-T^*(\omega(x_0))-1\})\\
    &\leq\tilde\beta(\omega(x_0),t+j).
\end{align*}
Hence we again conclude \eqref{case:combined}, which proves the claim. This completes the proof for the implication $3) \Rightarrow 2)$ in Theorem~\ref{thm:main}.\hfill$\blacksquare$
\end{pf*}



\section{Discussion and Conclusion}\label{sec:conclusion}
In this work we have first shown the equivalence between \russonew{0-UGAS} plus UBEBS and iISS for hybrid systems. In order to show this equivalence, some necessary Lyapunov-like conditions for \russonew{0-UGAS} and UBEBS are investigated. We also propose a non-smooth Lyapunov characterization. iISS is then shown when this theorem is applied to the sum of the two Lyapunov-like functions derived from \russonew{0-UGAS} and UBEBS. With the help of the aforementioned equivalence, the combination of local iISS and practical iISS, which are defined in this work, is also shown to be equivalent to iISS under one condition on the quantifiers.

\ifthenelse{\equal{\complete}{T}}{There are lots of extensions and further research directions that can be investigated. First of all, Lyapunov characterizations of local iISS and practical iISS can be studied. Inspired by the work~\cite{D2004Separation}, other asymptotic characterizations of iISS hybrid systems can also be examined. Meanwhile, by modeling switched systems as a particular type of hybrid systems and using our stability results for hybrid systems, it is possible to extend the results in our recent work~\cite{AR-SL-DL-AC:21} and conclude quantitative criteria on the switching frequency which guarantees the switched systems to be locally stable or practically stable. Similar approaches can also be used to study local or practical stability of sampled-data control systems or event-triggered systems. We can therefore design adaptive control strategies for such hybrid systems, which ensure global asymptotic stability with respect to the inputs by appealing to our theoretical results on the equivalent characterizations of stability notions that depend on input-to-state relations.}
{Many extensions and research directions can be explored, such as studying Lyapunov characterizations of local iISS and practical iISS. Examining asymptotic characterizations of iISS hybrid systems, inspired by~\cite{D2004Separation}, is another possibility. Furthermore, our stability results for hybrid systems can be used to extend the findings in our recent work~\cite{AR-SL-DL-AC:21} and establish quantitative criteria for the switching frequency ensuring local or practical stability of switched systems. 
}

\bibliographystyle{plain}

\end{document}